\documentclass[showpacs,preprintnumbers,amsmath,amssymb,pre,twocolumn]{revtex4}
\usepackage[dvips]{graphicx}
\usepackage{dcolumn}
\usepackage{amsmath}
\usepackage{amssymb}
\usepackage{bm}
\usepackage{xspace}

\RequirePackage{ifpdf}
\ifpdf
\usepackage[pdftex]{graphicx}
\else
\usepackage[dvips]{graphicx}
\fi

\newcommand{\nnn}{next-nearest neighbour\xspace}
\newcommand{\Neel}{N\'{e}el\xspace}

\newcommand{\Eqref}[1]{Eq.~\ref{#1}}

\newcommand{\Figref}[1]{Fig.~\ref{#1}}

\newcommand{\FigrefTo}[2]{Figs.~\ref{#1} and \ref{#2}}

\newcommand{\bvec}[1]{{\bf #1}}

\newcommand{\GnuplotRotatePS}{\ifpdf 0.0 \else -90 \fi}

\begin{document}

\title{Deconfined quantum criticality in the two dimensional Antiferromagnetic Heisenberg model with
  next nearest neighbour Ising exchange}
\author{J. Hove and A. Sudb{\o}}

\affiliation{%
  Department of Physics, Norwegian University of Science 
  and Technology, N-7491 Trondheim, Norway}
\date{\today}


\begin{abstract}
  We have considered the $S=1/2$ antiferromagnetic Heisenberg model in
  two dimensions, with an additional Ising \nnn interaction.
  Antiferromagnetic \nnn interactions will lead to frustration, and
  the system responds with flipping the spins down in the $xy$ plane.
  For large next nearest neighbour coupling the system will order in a
  striped phase along the $z$ axis, this phase is reached through a
  first order transition. We have considered two generalizations of
  this model, one with random \nnn interactions, and one with an
  enlarged unit cell, where only half of the atoms have \nnn
  interactions. In both cases the transition is softened to a second
  order transition separating two ordered states. In the latter case
  we have estimated the quantum critical exponent $\beta \approx
  0.25$. These two cases then represent candidate examples of
  deconfined quantum criticality.
\end{abstract}
\maketitle

\section{Introduction}
The groundstate of the antiferromagnetic Heisenberg model is
macroscopically degenerate. This makes it particularly sensitive to
additional interactions, which might induce transitions to different
states\cite{Cuccoli:2003}. In a seminal paper the concept of
fractionalized order, was set forth in Ref.  \cite{Senthil:2004}. The
generic starting point of this analysis is the two dimensional
antiferromagnetic Heisenberg model
\begin{equation}
  \label{Heisenberg}  
  H = J\sum_{\langle i,j\rangle} \vec{S}_i \cdot \vec{S}_j + \cdots,
\end{equation}
where the ellipsis represent additional short range interactions,
governed by a coupling $g$.  For $g = 0$ the groundstate is the
antiferromagnetic \Neel state, and by tuning $g$ the system can supposedly
be driven through a continuous quantum phase transition to a state
with a different type of order.  According to the
Landau-Ginzburg-Wilson (LGW) paradigm for phase transitions, an
order-to-order transition must either be first order, or 
proceed via an
intermediate disordered state. Since the scenario envisaged 
in Ref. \cite{Senthil:2004}, namely a {\it continuous} order-to-order transition, breaks with this
paradigm, the term ``deconfined criticality'' was coined to 
describe these transitions \cite{Senthil:2004}.

The only microscopic model considered in some detail in the context of
deconfined criticality is a dimer model with two spins in the unit
cell\cite{Sachdev:2004:book}. For this particular model the transition
between a \Neel state and a spin-gapped paramagnet can be shown
analytically, and are also confirmed with QMC
calculations\cite{Matsumoto:2001}. Apart from this dimer model it has
been difficult construct microscopic models which give rise to
deconfined criticality.
Sandvik et.al. have investigated a model with ferromagnetic XY
interactions and a four-spin ring
exhange\cite{Sandvik:2002,Melko:2004}, this model has a quantum
critical point separating a superfluid and valence bond solid, which
might be a microscopic manifestation of deconfined criticality.
Another possible way to build a microscopic model which might give
rise to deconfined criticality is to include frustration. A natural
way to frustrate the Heisenberg model is with a next nearest neighbour
(nnn) Heisenberg interaction; this is usually called the $J_1-J_2$
model.  The $J_2$ coupling will favor antiparallel spins along next
nearest neighbour bonds, this is in conflict with the nearest
neighbour exchange. The result is frustration, and a reduction in the
antiferromagnetic ordering.

Unfortunately, in this model the geometric frustration gives rise to a
sign-problem, and the model is really not amenable to a Monte Carlo
based approach.  Studies of this model have been based on a
reweighting technique\cite{Nakamura:1993}, exact
diagonalization\cite{Schulz:1996} and variational
methods\cite{Capriotti:2001}. The results indicate that \Neel order
persists up to $\kappa = J_2 / J_1 \lesssim 0.40$, and that a striped
order develops for $\kappa \gtrsim 0.60$. Recent results indicate that
the transition at $\kappa \sim 0.40$ is a weak first-order
transition\cite{Sirker:2006}.

To avoid the sign problem of the $J_1-J_2$ model, one can study 
a simplified model where the nnn exchange is only along the
$z$-components of the spin, i.e. the model
\begin{equation}
  \label{Eq:H}
  H =        J\big(\sum_{\langle i,j\rangle} \vec{S_i} \cdot \vec{S_j} + 
      \kappa \sum_{\langle\langle i,j\rangle\rangle} S_i^{z} S_j^{z}\big).
\end{equation}

This simplified model captures the effect of frustration, but in
contrast to the $J_1 - J_2$ model the isotropy in spin space is
explicitly broken by the \nnn interaction. Hence, in particular for
small $\kappa$ we expect this to be a much stronger perturbation of
the antiferromagnet Heisenberg model than the $J_2$ coupling.

Apart from the Heisenberg point at $\kappa = 0$ we expect three
different phases as $\kappa$ is varied: For $\kappa < 0$ the system is
not frustrated, and the additional \nnn will only serve to increase
the antiferromagnetic ordering. Observe however that the \nnn
interaction has singled out the $z$ direction in spin space, i.e.  the
model should be in the universality class of the Ising model and have
an ordered state at finite $T$. For $\kappa > 0$ the system will be
frustrated, for moderate $\kappa$ we expect that the system will avoid
the frustration by flipping the spins down in the $xy$ plane, i.e. we
will effectively get an antiferromagnetic $O(2)$ model. For larger
values of $\kappa$ the \nnn interaction will dominate, in which case
the spins will again point along the $z$ axis, and order in a state
with stripe order. The transition from the effective antiferromagnetic
$O(2)$ model to the striped state is first order. In Ref.
\cite{Roscilde:2004} an antiferromagnetic Heisenberg model with an
additional \emph{anisotropic} \nnn exchange was studied. This work
reported Monte Carlo results in the limit of zero transverse nnn
interactions; i.e. \Eqref{Eq:H}. In addition, a model similar to
\Eqref{Eq:H} in the terms of hardcore bosons was considered in Refs.
\cite{Hebert:2001,Schmid:2004}. We have deteremined the phase diagram
of \Eqref{Eq:H}, essentially reproducing the results of Refs.
\onlinecite{Roscilde:2004,Hebert:2001,Schmid:2004}. In addition, we
have considered two generalisations aimed at softening the first order
transition to the striped state.

The generalisations of \Eqref{Eq:H} we have considered are first a
disordered model, where the \nnn bond strength is $\kappa_0 \pm \Delta \kappa$ with equal probability. This is motivated from a theorem \cite{Aizenman:1989} which states that in two dimensions
any amount of bond disorder will be sufficient to soften a first order transition into a second order transition. Secondly, we have studied a model were only half of the sites have \nnn interaction. This model will clearly share many of the qualitative features of the original model, but the effect of the \nnn interactions will 
be reduced.

\section{QMC Simulations}
We have performed Quantum Monte Carlo simulations using the Stochastic Series Expansion (SSE)\cite{Sandvik:1991,Sandvik:1997} method. In the SSE method the Hamilton operator is written as 
a sum of bond operators
\begin{equation}
  \label{HBond}
  H = \sum_{b} \left(H_{d,b} + H_{od,b}\right).
\end{equation}
The sum in \Eqref{HBond} is over all the bonds on the lattice,
$H_{d,b}$ is an operator working on bond $b$, which is \emph{diagonal} in the basis chosen to represent the spin space, and $H_{od,b}$ is an off-diagonal operator. For spin models 
with $z$ axis magnetization as basis, the operator $H_{d,b}$ 
will be
\begin{equation}
  \label{Hd}
  H_{d,b} = J ~ S_{i(b)}^zS_{j(b)}^z,   
\end{equation}
where $i(b)$ and $j(b)$ are the two sites connected by bond $b$. $H_{od,b}$ is an off-diagonal operator, and in the case of spin
models we will have $H_{od,b}$ given by
\begin{equation}
  \label{Hod}
  H_{od,b} = \frac{J}{2} \left( S_{i(b)}^{+}S_{j(b)}^{-} + S_{i(b)}^{-}S_{j(b)}^{+}\right).
\end{equation}
Observe that for the actual simulations the operators $H_{d,b}$ 
are scaled and shifted\cite{Sandvik:1997} to ensure
\begin{align}
  \label{SSE:req}
  H_{d,b}  | \uparrow \downarrow \rangle &= | \uparrow \downarrow \rangle &   H_{d,b} | \uparrow \uparrow \rangle  &= 0.
\end{align}

The formal expression for the partition function is then expanded,
which yields the following representation
\begin{equation}
  \label{PartSSE}
  Z(\beta) = \sum_{\{\alpha\}} \sum_{n} \sum_{\{S_n\}} \frac{\left(-\beta\right)^n}{n!} \left\langle \alpha\left|\prod_{i}^n H_{\sigma_i}\right| \alpha \right\rangle.
\end{equation}
Here $S_n$ is a sequence of $n$ pairs, each pair consisting of 
a variable denoting operator type and a bond index, i.e.
\begin{equation}
  \label{Sn}
  S_{n} = \big\{\underbrace{\left(a_1,b_1\right)}_{\sigma_1}, \left(a_2,b_2\right),\ldots \left(a_n,b_n\right) \big\}.
\end{equation}
The variable $a_i$ in \Eqref{Sn} denotes \emph{type} of operator and
can be either diagonal or off-diagonal. The SSE method then consists
of doing importance sampling in the combined space $\left| \left\{
    \alpha \right\} \right \rangle \otimes S_n$. The actual updates
are of two different types. The diagonal updates insert or remove a
diagonal operator $H_{d,b}$, thereby changing the expansion order 
$n \to n \pm 1$. The off-diagonal operators change operator types $H_{d,b} \leftrightarrow H_{od,b}$ and flip the corresponding spins, this must be done in a way which ensures periodicity in the $\beta$ direction,
i.e. $|\alpha(0)\rangle = |\alpha(n)\rangle$. For the off diagonal
updates, the advent of loop updates \cite{Evertz:1993} has
significantly improved the performance of
SSE  simulations \cite{Sandvik:1999,Syljuaasen:2002}.

\begin{figure}[htbp]
  \centerline{\scalebox{0.50}{\rotatebox{0.00}{\includegraphics{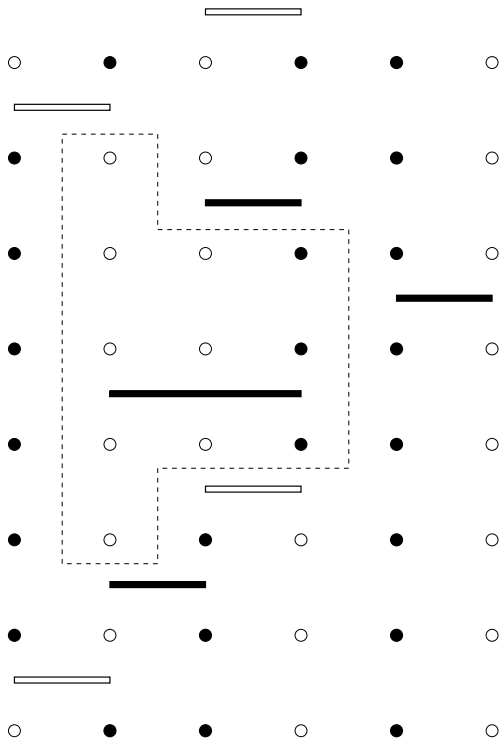}}}}
  \caption{\label{Freezing} A chain of six spins, depicted with 
  an operator sequence of length $n=8$ in the $\beta$ direction.
  The filled bars
    denote diagonal interactions $S_{i}^zS_j^z$ and the open bars are flip
    operators $S_i^{+}S_j^- + S_i^{-}S_j^+$. The dashed region shows spins/p-slices
    which have been \emph{frozen} by the \nnn bond in the middle.}
\end{figure}

For the ordinary $S=1/2$ Heisenberg model, SSE simulations with
operator loop update are particularly simple. In order to include the
\nnn interactions, we need to modify the algorithm slightly.  In the
case of the diagonal updates this merely amounts to including the
extra factor $\kappa$ in the weight calculation for the \nnn bonds.
Whereas for the operator loop the \nnn interactions have a more
profound effect. These interactions are \emph{only} diagonal, i.e. the
incoming and outgoing spin states must be equal. Furthermore, the \nnn
bonds can only connect antiparallel ($\kappa > 0$) spins. The result of
this is that the \nnn bonds ``freeze'' a substantial
part of the spin configuration, and only those spins/operators not
directly linked to a \nnn bond are amenable to operator loop update,
as illustrated in \Figref{Freezing}.  Clearly, this freezing affects
the performance of the simulations in a negative way, in particular
for intermediate values of $\kappa$.

\subsection{Observables}
To differentiate between the different types of order in the 
model, we have studied the structure factor
\begin{equation}
  \label{Sq}
  S(\bvec{Q}) = \frac{1}{N} \left\langle \bigg(\sum_{\bvec{r}} \langle \alpha | S^z(\bvec{r})| \alpha \rangle e^{i\bvec{Q} \cdot \bvec{r}} \bigg)^2\right\rangle,   
\end{equation}
for different values of $\bvec{Q}$.  An estimator of $S(\bvec{Q})$
taking all the intermediate SSE states into account can be found in
\cite{Sandvik:1997}.  For the remaining part of the text we will make
frequent use of the terms staggered and striped magnetization, these
quantities are defined as
\begin{align}
  \label{Mq}
  M^z_{(\pi,\pi)} &= \frac{1}{N}\sqrt{3S(\pi,\pi)},  \\
  \label{Mq_striped}
  M^z_{(\pi,0)} &= \frac{1}{N}\sqrt{S(\pi , 0 ) + S( 0, \pi )}. 
\end{align}
The upper index indicates that the magnetization is evaluated along
the $z$ axis, and the lower index is the direction of $\bvec{Q}$ 
in the evalution of \Eqref{Sq}, i.e. $(\pi,\pi)$ for staggered 
and $(\pi,0)$ for striped magnetization. The factor of three in \Eqref{Mq} is included to account for rotational averaging 
among the three directions in spin space. When $\kappa$ is 
finite, isotropy in spin space is explicitly broken.
We have nevertheless retained this factor to get continuous 
formulae around $\kappa = 0$. In addition to the structure 
factor, we have also measured two other quantities, namely 
specific heat and superfluid density. 

The specific heat is given by 
\begin{align}
  \label{SSEE}
  C_V &= \beta^2 \frac{\partial^2}{\partial \beta^2} \ln Z  = \langle n (n-1) \rangle - \langle n \rangle^2.
\end{align}
Here, $n$ refers to the summation 
variable in \Eqref{PartSSE}. This summation is truncated in a stochastic manner and  $n$ is thus promoted to a dynamical 
variable in SSE. We will not exhibit results for $C_V$ explicitly, but have used the anomalies in this quantity to corroborate the
phase boundaries shown in  \Figref{Fig:PhaseD} (with the
exception of the line separating the superfluid phase from
the disordered phase, see comments on this below).

The estimator for the superfluid density ($O(2)$ ordering) is 
given by \cite{Sandvik:1997}
\begin{equation}
  \label{rhos:estimator}
  \rho_S = \frac{3}{4\beta N}\left(\langle \left(N_x^+ - N_x^-\right)^2 \rangle + \langle \left(N_y^+ - N_y^-\right)^2 \rangle \right),
\end{equation}
where $N_{\mu}^+/N_{\mu}^-$ is the number of $S^+_iS^-_j$ 
and $S^-_iS^+_j$ operators applied along bonds in the 
$\mu$ direction.

\section{Results} 
\begin{figure}[htbp]
  \centerline{\scalebox{0.50}{\rotatebox{\GnuplotRotatePS}{\includegraphics{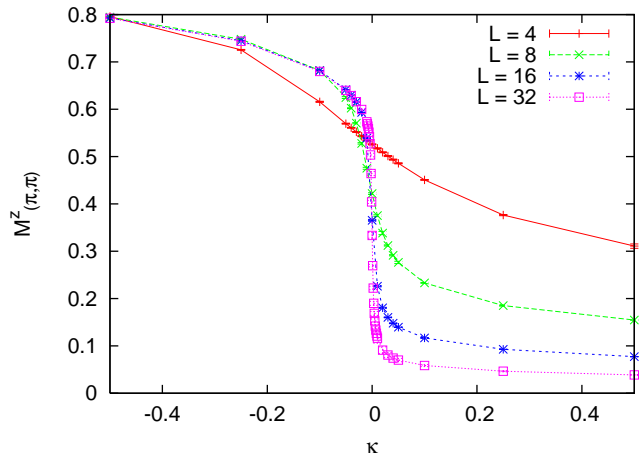}}}}
  \caption{\label{Mxy}(Color online) The staggered magnetization along the $z$ axis 
    in the ground state ($\beta = 10$), as a function of $\kappa$.}
\end{figure}

As mentioned previously, the model in the form of \Eqref{Eq:H} 
has already been studied in Ref. \cite{Roscilde:2004}. As a
benchmark of our QMC methods, we started with this model to reproduce the results of  Ref. \cite{Roscilde:2004}.  \FigrefTo{Mxy}{MxMy} shows the staggered and striped 
magnetization along the $z$ axis, as a function of $\kappa$.

\begin{figure}[htbp]
  \centerline{\scalebox{0.50}{\rotatebox{\GnuplotRotatePS}{\includegraphics{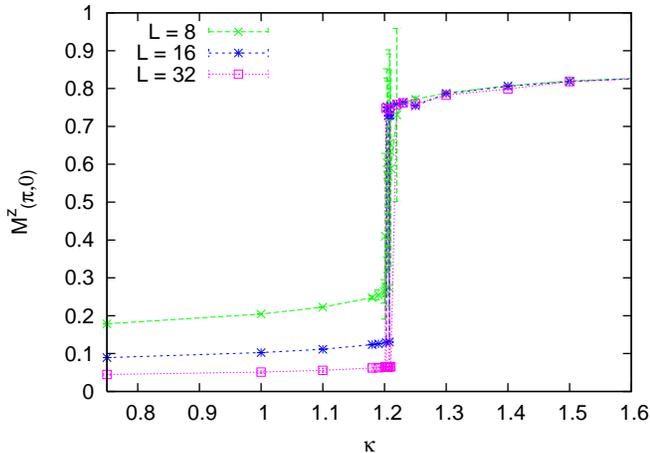}}}}
  \caption{\label{MxMy}(Color online) The striped magnetization in the ground state ($\beta = 10$), as a function of $\kappa$. 
    There is a first order transition at $\kappa \approx 1.205$.}
\end{figure}

From these two figures we conclude the following. (1) For negative
$\kappa$ the magnetization along the $z$ axis is enhanced; this 
is expected since the $\kappa < 0$ system is not frustrated. (2)
For $\kappa \gtrsim 0$ the magnetization is immediately tilted away from the $z$ axis, leaving zero magnetization along the $z$ axis. (3) For large $\kappa$ the magnetization again orders along the 
$z$ axis, in this case in a striped formation. The transition to the striped phase is a first order transition, the discontinuous jumps in $M^z(\pi,0)$ in \Figref{MxMy} indicate this, and it is also confirmed by a more detailed analysis of histograms of e.g. the striped order parameter\cite{Roscilde:2004} or number of \nnn 
bond operators. Hence, for this case the order-to-order transition depicted in \Figref{MxMy} (a transition from superfluid order,
equivalently $O(2)$ order, to stripe order) falls within the standard Landau-Ginzburg-Wilson paradigm of phase transitions.

\begin{figure}[htbp]
  \centerline{\scalebox{0.45}{\rotatebox{\GnuplotRotatePS}{\includegraphics{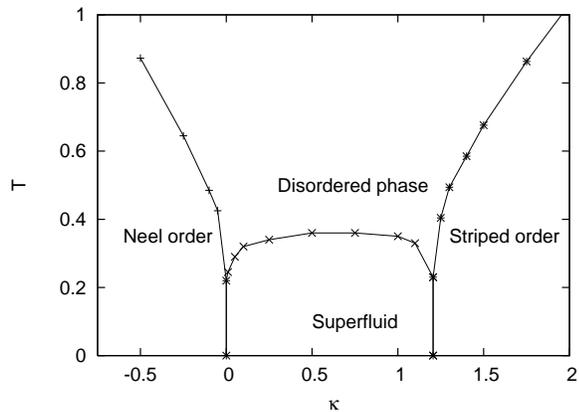}}}}
  \caption{\label{Fig:PhaseD} Phase diagram for the model in 
  the $\kappa,T$  plane. For $\kappa > \kappa_c$, and 
  $\kappa < 0$ the model has magnetic order, striped and 
  staggered respectively. In the intermediate $\kappa$ 
  range values there is no finite $T$ magnetic order, 
  however there is a superfluid order which persists into 
  the finite $T$ region. The line separating
  topological $2DXY$ order from the normal phase is determined 
  with considerably less precision than the two other lines.} 
\end{figure}

For $\kappa < 0$ and $\kappa > 0$ the ordered state breaks a discrete symmetry, and the order persists for finite $T$. In 
the intermediate regime, $0 < \kappa \lesssim 1.205$ the 
remaining model is an antiferromagnetic $2DXY$ model with a continuous $O(2)$ symmetry. According to the Mermin Wagner 
theorem this symmetry can not  be spontaneously broken at 
finite $T$. However, there is finite  spin stiffness and topological order up to a temperature $T_{\mathrm{BKT}}$ where the order vanishes in a \emph{Berezinski-Kosterlitz-Thouless} transition.
The boundary of this region has been (approximately) 
located by equating $\rho_S(T)$ with $2T/\pi$. All in 
all we have found the phase diagram presented in \Figref{Fig:PhaseD} for \Eqref{Eq:H}.

\subsection{Disordered system}
\label{DisorderB}
\begin{figure}[htbp]
  \centerline{\scalebox{0.50}{\rotatebox{\GnuplotRotatePS}{\includegraphics{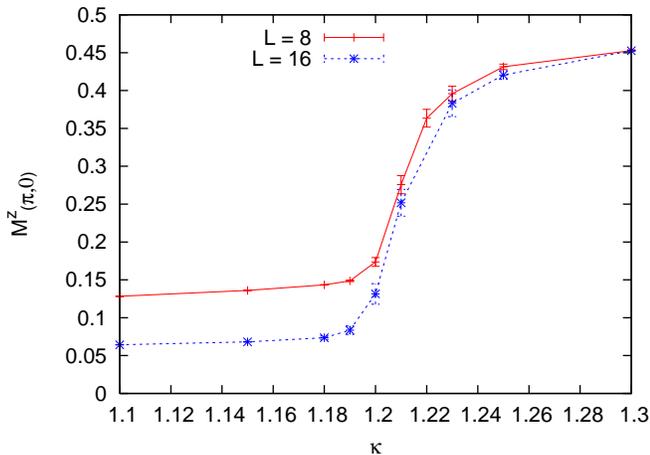}}}}
  \caption{\label{Disorder:MxMy}(Color online) Disorder 
  averaged value of $M^z(\pi,0)$ as a function of $\kappa_0$ for two different system sizes. The 
  indicated error bars are standard error estimates from the 
  independent disorder realizations. The temperature coupling 
  is $\beta = 10$.}
\end{figure}

The effect of disorder on phase transitions is a much studied 
topic. In the case of continuous transitions, the Harris
criterion \cite{Harris:1974,Chayes:1986} states that disorder 
will change the universality class of the transition, i.e. 
be relevant, if the exponents of the pure system satisfy 
$\nu < 2/d$. In the case of first order transitions, disorder 
can soften the transition into a continuous transition, in 
the case of two dimensions any amount of disorder is 
sufficient \cite{Aizenman:1989}, whereas a finite amount is
needed in three dimensions. These predictions have been 
confirmed for the Potts model in both two and three dimension\cite{Chen:1995,Chatelain:2005}. We have, 
however, not found tests of these predictions for a first 
order \emph{quantum} phase transition. We have investigated 
what happens with the first order quantum phase transition 
at $\kappa \approx 1.205$ when disorder is included in the
model. Along each bond is $\kappa$ is given by
\begin{equation}
  \label{kappaDisordered}
  \kappa = \kappa_0 \pm \Delta \kappa,
\end{equation}
with equal probability. We have focused on the striped 
order parameter $M^z(\pi,0)$ in the vicinity of 
$\kappa \approx 1.205$, in order to compare with 
\Figref{MxMy}. In the disordered system we must perform 
both ordinary thermodynamic averaging and subsequently 
disorder averaging. In e.g.  \Figref{Disorder:MxMy} the 
plotted 
quantity is given by

\begin{equation}
  \label{MxMy:D}
  \overline{M^{z}(\pi,0)} = \frac{1}{N}\sum_{i = 1}^N M^{z}_i(\pi,0), 
\end{equation}
where $M^z_i(\pi,0)$ is the striped magnetization in disorder
realization $i$, calculated according to \Eqref{Mq_striped}, and $N$
is the total number of disorder realizations. The number of disorder
realizations has typically been $N = 100$.

\Figref{Disorder:MxMy} shows the disorder averaged striped
magnetization as a function of $\kappa_0$ for $\Delta \kappa = 0.05$.
The strongly discontinuous features of $M^z(\pi,0)$ from \Figref{MxMy}
are washed out when disorder is introduced. From this, we conclude
that the transition changes order when disorder is introduced. The
location of the critical point coincides with the original transition
point of the uniform system. We have not varied $\Delta \kappa$
systematically, our results (not shown) indicate that the system
is not very sensitive to $\Delta \kappa$ variations.

The low-$\kappa$ region in \Figref{Disorder:MxMy} is an $O(2)$ ordered
state. The large-$\kappa$ region is an Ising-ordered state with an
additional stripe order.  {\it Hence, this quantum phase transition is
  a transition from an ordered state to another ordered state, and
  since it is continuous, it represents a candidate example of
  so-called deconfined quantum criticality.} 

\subsection{AB system}
\begin{figure}[htbp]
  \centerline{\scalebox{0.35}{\rotatebox{0.00}{\includegraphics{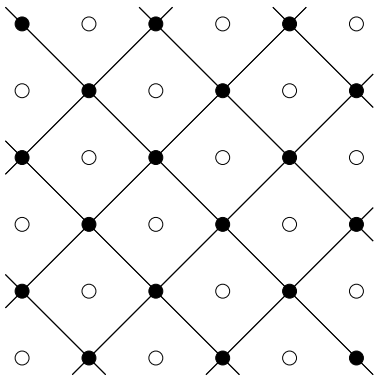}}}}
  \caption{\label{ABLattice} The binary system consisting of 
  two different atomic species. Only type A $(\bullet)$ has 
  \nnn interaction, illustrated with diagonal lines.}
\end{figure}

For large $\kappa$ the Ising \nnn interaction in \Eqref{Eq:H} is 
a very strong interaction. In an attempt to soften the transition to the high $\kappa$ state into a continuous transition we have devised a model consisting of two ``atom'' types A and B, where 
the Ising \nnn interaction is only between the $A$ atoms, the scenario is illustrated
in \Figref{ABLattice}.

\begin{figure}[htbp]
  \centerline{\scalebox{0.50}{\rotatebox{\GnuplotRotatePS}{\includegraphics{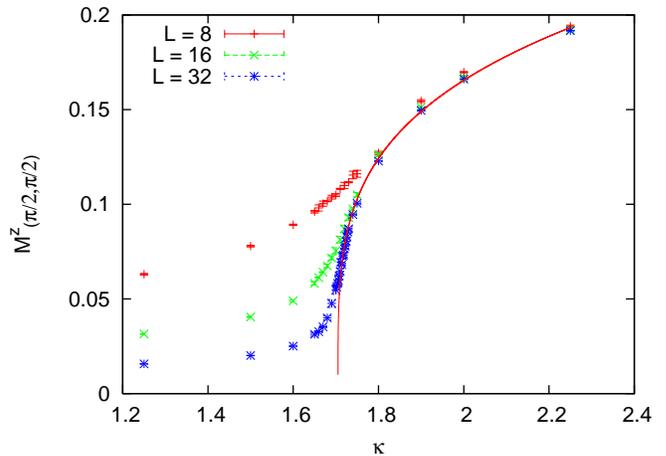}}}}
  \caption{\label{MAB}(Color online) The staggered magnetization 
  of the $A$ sites in the groundstate of the AB model, as a
  function of $\kappa$. The solid line is a least squares 
  fit to a power law.}
\end{figure}

This model has many of the same qualitative properties as the original model, in particular small values of $\kappa$ will frustrate the system and tilt the magnetization down in the 
$xy$ plane. For large $\kappa$ the A atoms will form an antiferromagnetically ordered state, with AF magnetization 
along the $z$ axis. In this state the A and B sites will 
decouple, and the B sites will be disordered with no net
contribution to the energy of the system. As an order 
parameter for this transition we have considered the 
staggered magnetization along the $z$ axis, for the 
$A$ sites, i.e.
\begin{equation}
  \label{Eq:MAB}
  M^{z}_{(\pi/2,\pi/2)} = \frac{1}{N}\sqrt{S_{\mathrm{A}}(\pi/2,\pi/2)},
\end{equation}
where the sum is only over $A$ sites. Because the sum is limited 
to the $A$ sites, full polarization corresponds to $M^z = 0.25$. \Figref{MAB} shows the staggered magnetization among the 
$A$ sites as a function of $\kappa$.

Comparing with the striped magnetization of the uniform model,
\Figref{MxMy}, we see that the critical coupling $\kappa_c \approx
1.705$ of the AB system is much larger than for the uniform system.
This is reasonable, since the \nnn interaction only operates on half
of the sites. Furthermore, \Figref{MAB} shows that the transition to
the ordered state is much \emph{smoother} than in the uniform model,
hence the figure indicates that the transition is continuous. The
continuous nature of the transition is also confirmed by considering
the number distribution of e.g. \nnn operators at critical point. This
quantity is unimodal, whereas for the uniform model it is strongly
bimodal, reminiscent of a first order transition. From this, we
conclude that the AB-model deformation of the Hamiltonian in
\Eqref{Eq:H} suffices to promote quantum deconfined criticality.
Stripe order is lost by removing every second next-nearest neighbor
coupling, but an order-to-order quantum phase transition nevertheless
remains. Namely, the transition is that from an $O(2)$ (superfluid)
ordered state at low and intermediate values of $\kappa$ to a $Z_2$
(Ising) ordered state at high values of $\kappa$. {\it Since the
  transition is second order, it represents a second candidate example
  of deconfined quantum criticality.} 

We have not made attempts at completely determining the critical
exponents at the transition. However a fit of $M^z$ from the 
$L=32$ system to the functional form
\begin{equation}
  \label{MAB:Fit}
  M^{z}_{(\pi/2,\pi/2)} \propto | \kappa - \kappa_c |^{\beta}  
\end{equation} 
with $\kappa_c = 1.705$ gave good results, with a critical 
exponent $\beta \approx 0.25$. The fit is shown as a solid 
line in \Figref{MAB}.

\section{Conclusion}
We have studied varieties of the antiferromagnetic Heisenberg model in
two dimensions with additional \nnn Ising exchange. This model has a
strong first order transition at $\kappa_c \approx 1.205$.  We have
studied two generalisations of the model, one based on disordered \nnn
couplings, and another where only half the sites are endowed with \nnn
interaction. Both the generalised models feature continuous quantum
phase transitions from one ordered state to another ordered state.
{\it As such, these two cases represent candidate examples of
  deconfined quantum criticality.} Frustrated interactions is an
essential part of the models we have considered, and as such they are
distinct from the model already considered by Sandvik et.al.
\cite{Sandvik:2002,Melko:2004}, where (possible) deconfined
criticality is brought about by ring-exchange.

\section{Acknowledgement}
This work was supported in part by the Research Council of Norway
through Grants No. 157798/432 and 158547/431 (NANOMAT) and
167498/V30(STORFORSK). Bergen Center for Computational Science (BCCS) is acknowledged for computing time. 


\end{document}